\documentstyle[preprint,aps]{revtex}
\tightenlines
\begin{document}
\draft
\title{Fractal sets of dual topological quantum numbers}

\author{Wellington da Cruz}
\address{Departamento de F\'{\i}sica,\\
 Universidade Estadual de Londrina, Caixa Postal 6001,\\
Cep 86051-970 Londrina, PR, Brazil\\
E-mail address: wdacruz@exatas.uel.br}
\date{\today}
\maketitle
\begin{abstract}

The universality classes of the quantum Hall transitions are 
considered in terms of fractal sets of dual topological quantum numbers 
filling factors, labelled by a fractal or Hausdorff dimension
defined into the interval $1 < h < 2$ and associated with fractal curves. 
We show that our approach to the fractional quantum Hall effect-FQHE is free 
of any empirical formula and this characteristic appears as a crucial insight 
for our understanding of the FQHE. According to our formulation, 
the FQHE gets a fractal structure from the connection between 
the filling factors and the Hausdoff dimension of the quantum paths 
of particles termed fractons which obey a fractal distribution 
function associated with a fractal von Neumann entropy. 
This way, the quantum Hall transitions satisfy some properties 
related to the Farey sequences of rational numbers and so our theoretical 
description of the FQHE establishes a connection between physics, 
fractal geometry and number theory. The FQHE as a convenient 
physical system for a possible prove of the Riemann hypothesis is suggested.

\end{abstract}

\pacs{PACS numbers: 71.10.Pm, 05.30.-d, 73.43.Cd, 05.30.Pr\\
Keywords:  Fractal distribution function; Fractal von Neumann entropy; 
Fractons; Fractional quantum Hall effect.}

\newpage

\section{Introduction}

The mathematical ideas about fractal geometry\cite{R1} have been applied in 
various contexts of physics, as in nonlinear and 
nonequilibrium phenomena\cite{R2}. Here, in this Chapter, we show that the 
fractional quantum Hall effect-FQHE\cite{R3} 
has a fractal-like structure such that the universality classes of the quantum Hall 
transitions constitute fractal sets labelled by the Hausdorff dimension 
defined within the interval $1$$\;$$ < $$\;$$h$$\;$$ <$$\;$$ 2$ and associated with 
fractal curves ( continuous functions but nowhere differentiable ) of 
objects called fractons. These ones are charge-flux systems defined 
in two-dimensional multiply connected space and they carry 
rational or irrational values of spin. The topological meaning 
of the dual filling factors which characterizes the FQHE comes 
from the connection between the fractal dimension of the 
quantum paths of fractons and these quantum numbers. Thus, we have 
introduced {\it a new geometric insight} for understanding that phenomenon.

Topological quantum numbers are insensitive to the imperfections 
of the systems and so the {\it fractal dimension} as a topological invariant 
makes robust the FQHE properties. By contrast, in the literature, some 
authors have considered 
the concepts of Chern numbers and 
Fredholm indices for the integer quantum Hall effect, but for the FQHE 
this research line is an open challenge. The Fredhom index is related to the creation and 
annihilation operators of many-body quantum mechanics and 
in some cases coincides with the Chern number\cite{R4}. The mathematical 
mechanism responsible for the FQHE has been also considered using some techniques of 
noncommutative geometry in connection with twisted higher index theory of elliptic 
operators on orbifold ( its definition generalizes the idea of a manifold ) 
covering spaces of compact good orbifolds, 
where the topological nature of the Hall condutance 
is stable under small deformations of a Hamiltonian, with the interaction 
simulates by the negative curvature of the hyperbolic structure of the model. 
The twisted 
higher index can be a fraction when the orbifold is not smooth and 
the topological invariant is namely the orbifold Euler characteristic\cite{R5}. The 
connection between the FQHE and topological Chern-Simons field theories 
is another route of investigation\cite{R6}.  
Therefore, our fractal approach offers 
{\it a possible alternative} for understanding some deeper 
mathematical and physical features underlying the FQHE
\cite{R7,R8,R9,R10,R11,R12,R13,R14,R15}. 

Fractons satisfy a fractal distribution 
function associated with a fractal von Neumann entropy and they are 
classified in universal classes of particles or quasiparticles. Our 
formulation was introduced in\cite{R7} 
and in particular, we have found an expression which relates 
the fractal dimension $h$ and the spin $s$ of the particles, 
$h=2-2s$, $0 < s < \frac{1}{2}$. This result is  
analogous to the fractal dimension formula 
of the graph of the functions, in the context of the fractal geometry, 
and given by: $\Delta(\Gamma)=2-H$, where $H$ is 
known as H\"older exponent, with $0 < H < 1$\cite{R1}. 
The bounds of the fractal dimension $1$$\;$$ < $$\;$$\Delta(\Gamma)$$\;$$ 
<$$\;$$ 2$ needs to be obeyed in order for a function 
to be a fractal, so the bounds of our parameter $h$ are 
defined such that, for $h=1$ 
we have fermions, for $h=2$ we have bosons and for 
$1$$\;$$ < $$\;$$h$$\;$$ <$$\;$$ 2$ we have fractons. The H\"older exponent 
characterizes irregular functions which appears in diverse physical systems. 
For instance, in the Feynman path integral approach of the quantum mechanics, 
the fractal character of the quantum paths was just observed\cite{R2}.   

The fractal properties of the 
quantum paths can be extracted from the propagators of the particles 
in the momentum space\cite{R7,R16} and so our expression relating 
$h$ and $s$ can once more be justified. On the other hand, the physical formula 
introduced by us, when we consider 
the spin-statistics relation $\nu=2s$, is written as $h=2-\nu$, $0 < \nu < 1$. 
This way, a fractal spectrum was defined taking into account a mirror symmetry:

\begin{eqnarray}
h-1&=&1-\nu,\;\;\;\; 0 < \nu < 1;\;\;\;\;\;\;\;\;h-1=
\nu-1,\;\;\;\;\;\;\; 1 <\nu < 2;\nonumber\\
h-1&=&3-\nu,\;\;\;\; 2 < \nu < 3;\;\;\;\;\;\;\;\;
 h-1=\nu-3,\;
\;\;\;\;\;\; 3 <\nu < 4;\;etc.
\end{eqnarray}

The statistical weight for these classes of fractons is given by\cite{R7}

\begin{equation}
\label{e11}
{\cal W}[h,n]=\frac{\left[G+(nG-1)(h-1)\right]!}{[nG]!
\left[G+(nG-1)(h-1)-nG\right]!}
\end{equation}

and from the condition of the entropy be a maximum, we obtain 
the fractal distribution function

\begin{eqnarray}
\label{e.44} 
n[h]=\frac{1}{{\cal{Y}}[\xi]-h}.
\end{eqnarray}

The function ${\cal{Y}}[\xi]$ satisfies the equation 

\begin{eqnarray}
\label{e.4} 
\xi=\biggl\{{\cal{Y}}[\xi]-1\biggr\}^{h-1}
\biggl\{{\cal{Y}}[\xi]-2\biggr\}^{2-h},
\end{eqnarray}

\noindent with $\xi=\exp\left\{(\epsilon-\mu)/KT\right\}$. We understand the 
fractal distribution function as a quantum-geometrical 
description of the statistical laws of nature, 
since the quantum path is a fractal curve as noted by Feynman and this 
reflects the Heisenberg uncertainty principle. The Eq.(\ref{e.44}) 
embodies nicely this subtle information about the quantum paths 
associated with the particles. 

We can obtain for any class its distribution function considering the
Eqs.(\ref{e.44},\ref{e.4}). For example, 
the universal class $h=\frac{3}{2}$ with distinct values of spin 
$\biggl\{\frac{1}{4},\frac{3}{4},\frac{5}{4},\cdots\biggr\}_{h=\frac{3}{2}}$, 
has a specific fractal distribution

\begin{eqnarray}
n\left[\frac{3}{2}\right]=\frac{1}{\sqrt{\frac{1}{4}+\xi^2}}.
\end{eqnarray}

\noindent This result coincides with another 
one of the literature of fractional spin particles for 
the statistical parameter $\nu=\frac{1}{2}$\cite{R17}, 
however our interpretation is completely distinct. This particular example, 
shows us that the fractal distribution is the same 
for all the particles into the universal class labelled by $h$ and with 
different values of spin.  Thus, we emphasize 
that in our formulation the spin-statistics 
connection is valid for such fractons. The authors in\cite{R17} 
never make mention to this possibility. Therefore, 
our results give another perspective for the fractional 
spin particles or anyons\cite{R6}. On the other hand, we can obtain straightforward the 
Hausdorff dimension associated to the quantum paths of 
the particles with any value of spin.  

We also have
 
\begin{eqnarray}
\xi^{-1}=\biggl\{\Theta[{\cal{Y}}]\biggr\}^{h-2}-
\biggl\{\Theta[{\cal{Y}}]\biggr\}^{h-1},
\end{eqnarray}

\noindent where

\begin{eqnarray}
\Theta[{\cal{Y}}]=
\frac{{\cal{Y}}[\xi]-2}{{\cal{Y}}[\xi]-1}
\end{eqnarray}

\noindent is the single-particle partition function. 
We verify that the classes $h$ satisfy a duality symmetry defined by 
${\tilde{h}}=3-h$. So, fermions and bosons come as dual particles. 
As a consequence, we extract a fractal 
supersymmetry which defines pairs of particles $\left(s,s+\frac{1}{2}\right)$. 
This way, the fractal distribution function appears as 
a natural generalization of the fermionic and bosonic 
distributions for particles with braiding properties. Therefore, 
our approach is a unified formulation 
in terms of the statistics which each universal class of 
particles satisfies, from a unique expression 
we can take out any distribution function. In some sense , we can say that 
fermions are fractons of the class $h=1$ and  
bosons are fractons of the class $h=2$.

The free energy for particles in a given quantum state is expressed as

\begin{eqnarray}
{\cal{F}}[h]=KT\ln\Theta[{\cal{Y}}].
\end{eqnarray}

\noindent Hence, we find the average occupation number

\begin{eqnarray}
\label{e.h} 
n[h]&=&\xi\frac{\partial}{\partial{\xi}}\ln\Theta[{\cal{Y}}].
\end{eqnarray}

\noindent The fractal von Neumann entropy per state in terms of the 
average occupation number is given as\cite{R7,R8} 

\begin{eqnarray}
\label{e5}
{\cal{S}}_{G}[h,n]&=& K\left[\left[1+(h-1)n\right]\ln\left\{\frac{1+(h-1)n}{n}\right\}
-\left[1+(h-2)n\right]\ln\left\{\frac{1+(h-2)n}{n}\right\}\right]
\end{eqnarray}

\noindent and it is associated with the fractal distribution function Eq.(\ref{e.44}).

Now, as we can check, each universal class $h$ of particles, 
within the interval of definition has its entropy defined 
by the Eq.(\ref{e5}). Thus, for fractons of the self-dual class
$\biggl\{\frac{1}{4},
\frac{3}{4},\frac{5}{4},\cdots\biggr\}_{h=\frac{3}{2}}$, we obtain
  
\begin{eqnarray}
{\cal{S}}_{G}\left[\frac{3}{2}\right]=K\left\{(2+n)\ln\sqrt{\frac{2+n}{2n}}
-(2-n)\ln\sqrt{\frac{2-n}{2n}}\right\}. 
\end{eqnarray}

\noindent The Fermi and Bose statistics ( and the respective entropies, free energies ) associated to 
the universal classes of the fermions ($h=1$) and bosons ($h=2$) are recovered promptly.

We have also introduced the topological concept of fractal index, 
which is associated with each class. As we saw, $h$ is a geometrical parameter 
related to the quantum paths of the particles and so, we define\cite{R9} 

\begin{equation}
\label{e.1}
i_{f}[h]=\frac{6}{\pi^2}\int_{\infty(T=0)}^{1(T=\infty)}
\frac{d\xi}{\xi}\ln\left\{\Theta[\cal{Y}(\xi)]\right\}.
\end{equation}

\noindent For the interval of the definition $ 1$$\;$$ \leq $$\;$$h$$\;$$ \leq $$\;$$ 2$, there 
exists the correspondence $0.5$$\;$$ 
\leq $$\;$$i_{f}[h]$$\;$$ \leq $$\;$$ 1$, which signalizes 
a connection between fractons and quasiparticles of the conformal field theories, 
in accordance with the unitary $c$$\;$$ <$$\;$$ 1$ 
representations of the central charge. For $\nu$ even it is defined by 

\begin{eqnarray}
\label{e.11}
c[\nu]=i_{f}[h,\nu]-i_{f}\left[h,\frac{1}{\nu}\right]
\end{eqnarray}

\noindent and for $\nu$ odd it is defined by 

\begin{eqnarray}
\label{e.12}
c[\nu]=2\times i_{f}[h,\nu]-i_{f}\left[h,\frac{1}{\nu}\right],
\end{eqnarray}

\noindent where $i_{f}[h,\nu]$ means the fractal 
index of the universal class $h$ which contains the particles 
with distinct values of spin. 
In another way, the central charge $c[\nu]$ can be obtained using the 
Rogers dilogarithm function, i.e. 

\begin{equation}
\label{e.16}
c[\nu]=\frac{L[x^{\nu}]}{L[1]},
\end{equation}

\noindent with $x^{\nu}=1-x$,$\;$ $\nu=0,1,2,3,etc.$ and 

\begin{equation}
L[x]=-\frac{1}{2}\int_{0}^{x}\left\{\frac{\ln(1-y)}{y}
+\frac{\ln y}{1-y}\right\}dy,\; 0 < x < 1.
\end{equation}

\noindent Thus, we have established a connection between fractal geometry and 
number theory, given that the dilogarithm function appears 
in this context, besides another branches of mathematics\cite{R18}. 

Such ideas can be applied in the context of the FQHE. This phenomenon is 
characterized  by the filling factor parameter $f$, and for 
each value of $f$ we have the 
quantization of Hall resistance and a superconducting state 
along the longitudinal direction of a planar system of electrons, which are
manifested by semiconductor doped materials, i.e., heterojunctions 
under intense perpendicular magnetic fields and lower 
temperatures\cite{R3}. 

The parameter $f$ is defined by $f=N\frac{\phi_{0}}{\phi}$, where 
$N$ is the electron number, 
$\phi_{0}$ is the quantum unit of flux and
$\phi$ is the flux of the external magnetic field throughout the sample. 
The spin-statistics relation is given by 
$\nu=2s=2\frac{\phi\prime}{\phi_{0}}$, where 
$\phi\prime$  is the flux associated with the charge-flux 
system which defines the fracton $(h,\nu)$. According to our approach 
there is a correspondence between $f$ and $\nu$, numerically $f=\nu$. 
This way, we verify that the filling factors 
experimentally observed  appear into the classes $h$ and from the definition of duality 
between the equivalence classes, we note that the FQHE occurs in pairs 
 of these dual topological quantum numbers.

\section{Fractal approach to the FQHE}

In\cite{R14,R15} we have considered recent experimental results 
and checked the validity of our fractal formulation to the FQHE. 
Another fractal approach to the FQHE was considered in\cite{R19}. We show that 
our formulation is free of any empirical formula, i.e., 
our theoretical description of the FQHE has a foundational 
basis connecting a fractal 
parameter $h$ associated with the quantum paths and the spin $s$ of fractons. 
Besides this, these particles obey a fractal distribution 
function associated with a fractal von Neumann entropy, 
as we discussed earlier. On the other hand, 
the suggestion in\cite{R20} about {\it the fractal-like 
structure to a deeper understanding of FQHE} was antecipated 
by us\cite{R7,R8,R9,R10,R11,R12,R13}, using 
properly concepts of the fractal geometry.

Let us consider the filling factors suggested in\cite{R20,R21,R22}  
and the FQHE data\cite{R19,R23,R24,R25,R26,R27} together. Taking into account 
the fractal spectrum and the duality symmetry, we can write the sequence:

\begin{eqnarray}
&&{\bf \frac{4}{1}}\;\rightarrow\frac{19}{5}\rightarrow\;
\frac{34}{9}\rightarrow\frac{15}{4}\rightarrow\frac{26}{7}\rightarrow
\frac{11}{3}\rightarrow\;
\frac{18}{5}\rightarrow\frac{25}{7}\rightarrow
\frac{7}{2}\rightarrow\frac{24}{7}\rightarrow
\frac{17}{5}\rightarrow\frac{10}{3}\rightarrow\;\;
\frac{23}{7}\rightarrow\nonumber\\
&&\frac{13}{4}\rightarrow
\frac{29}{9}\rightarrow\frac{16}{5}\rightarrow\;
{\bf \frac{3}{1}}\rightarrow\;\frac{14}{5}\rightarrow
\frac{25}{9}\rightarrow\;\frac{11}{4}\rightarrow
\frac{19}{7}\rightarrow\frac{8}{3}\rightarrow
\frac{13}{5}\rightarrow\frac{18}{7}\rightarrow\;
\frac{23}{9}\rightarrow\;\frac{28}{11}
\rightarrow\nonumber\\
&&\frac{33}{13}\rightarrow\;
\frac{5}{2}\rightarrow\;\frac{32}{13}\rightarrow
\frac{27}{11}\rightarrow\frac{22}{9}\rightarrow
\frac{17}{7}\rightarrow\;\frac{12}{5}\rightarrow\;
\frac{7}{3}\rightarrow\frac{16}{7}\rightarrow\;\frac{9}{4}
\rightarrow\;\frac{20}{9}\rightarrow
\frac{11}{5}
\rightarrow\;\;{\bf \frac{2}{1}}\rightarrow\nonumber\\
&&\frac{15}{8}
\rightarrow\frac{43}{23}\rightarrow\frac{28}{15}\rightarrow
\frac{13}{7}\rightarrow\;\frac{11}{6}\rightarrow
\frac{9}{5}\rightarrow\;\;\frac{16}{9}\rightarrow\;
\frac{7}{4}\rightarrow\frac{19}{11}\rightarrow
\frac{12}{7}\rightarrow\frac{17}{10}\rightarrow\;
\frac{5}{3}\rightarrow\;\;\;\frac{8}{5}\rightarrow\nonumber\\
&&\frac{11}{7}\rightarrow\frac{14}{9}\rightarrow
\frac{17}{11}\rightarrow\frac{20}{13}\rightarrow\;
\frac{3}{2}\rightarrow\;\frac{19}{13}\rightarrow\;
\frac{16}{11}\rightarrow\frac{13}{9}\rightarrow
\frac{10}{7}\rightarrow\frac{7}{5}\rightarrow\;\;
\frac{4}{3}\rightarrow\;\frac{13}{10}\rightarrow\;\;
\frac{9}{7}\rightarrow\nonumber\\
&&\frac{14}{11}\rightarrow\;
\frac{5}{4}\rightarrow\;\frac{11}{9}\rightarrow\;
\frac{6}{5}\rightarrow\;\;\frac{7}{6}\rightarrow\;
\frac{8}{7}\rightarrow\;\;\frac{17}{15}\rightarrow
\frac{26}{23}\rightarrow\;\frac{9}{8}\rightarrow\;
{\bf \frac{1}{1}}\;\;\rightarrow\frac{8}{9}\rightarrow\;\;
\frac{15}{17}\rightarrow\;
\frac{7}{8}\rightarrow\nonumber\\
&&\frac{13}{15}\rightarrow\;
\frac{6}{7}\rightarrow\;\frac{11}{13}\rightarrow
\frac{16}{19}\rightarrow\;\frac{5}{6}\rightarrow
\frac{19}{23}\rightarrow\;\;\frac{14}{17}\rightarrow
\frac{9}{11}\rightarrow\frac{22}{27}\rightarrow
\frac{13}{16}\rightarrow\frac{17}{21}\rightarrow
\frac{21}{26}\rightarrow\;
\frac{25}{31}\rightarrow\nonumber\\
&&\frac{4}{5}\rightarrow\;\frac{23}{29}\rightarrow\;
\frac{19}{24}\rightarrow\frac{15}{19}\rightarrow
\frac{11}{14}\rightarrow\frac{18}{23}\rightarrow\;\;
\frac{7}{9}\rightarrow\;\frac{10}{13}\rightarrow
\frac{13}{17}\rightarrow\frac{16}{21}\rightarrow
\frac{19}{25}\rightarrow\;\frac{3}{4}\rightarrow\;
\frac{17}{23}\rightarrow\nonumber\\
&&\frac{14}{19}\rightarrow\frac{11}{15}\rightarrow
\frac{8}{11}\rightarrow\;\frac{5}{7}\rightarrow\;
\frac{12}{17}\rightarrow\frac{7}{10}\rightarrow\;
\frac{16}{23}\rightarrow\frac{9}{13}\rightarrow
\frac{20}{29}\rightarrow\frac{11}{16}\rightarrow
\frac{24}{35}\rightarrow\frac{13}{19}\rightarrow\;
\frac{2}{3}\rightarrow\\
&&\frac{11}{17}\rightarrow
\frac{20}{31}\rightarrow\frac{9}{14}\rightarrow
\frac{16}{25}\rightarrow\frac{7}{11}\rightarrow
\frac{12}{19}\rightarrow\;\frac{17}{27}\rightarrow\;
\frac{5}{8}\rightarrow\;\frac{18}{29}\rightarrow\;
\frac{13}{21}\rightarrow\frac{8}{13}\rightarrow
\frac{3}{5}\;\rightarrow\frac{16}{27}\rightarrow\nonumber\\
&&\frac{29}{49}\rightarrow
\frac{13}{22}\rightarrow\frac{23}{39}\rightarrow
\frac{10}{17}\rightarrow\frac{17}{29}\rightarrow
\frac{24}{41}\rightarrow\;\frac{7}{12}\rightarrow
\frac{25}{43}\rightarrow\frac{18}{31}\rightarrow\;
\frac{11}{19}\rightarrow\;\frac{4}{7}\rightarrow
\frac{21}{37}\rightarrow\frac{38}{67}\rightarrow\nonumber\\
&&\frac{17}{30}\rightarrow\frac{30}{53}\rightarrow
\frac{13}{23}\rightarrow\frac{22}{39}\rightarrow
\frac{31}{55}\rightarrow\frac{9}{16}\rightarrow\;
\frac{32}{57}\rightarrow\frac{23}{41}\rightarrow
\frac{14}{25}\rightarrow\;\frac{5}{9}\rightarrow
\;\frac{6}{11}\rightarrow\frac{7}{13}\rightarrow
\frac{8}{15}\rightarrow\nonumber\\
&&\frac{9}{17}\rightarrow\frac{10}{19}\rightarrow
\frac{11}{21}\rightarrow\frac{1}{2}\rightarrow\;\;
\frac{10}{21}\rightarrow\frac{9}{19}\rightarrow
\frac{8}{17}\rightarrow\frac{7}{15}\rightarrow
\frac{6}{13}\rightarrow\;\frac{5}{11}\rightarrow\;
\frac{4}{9}\rightarrow\;\frac{11}{25}\rightarrow
\frac{18}{41}\rightarrow\nonumber\\
&&\frac{25}{57}\rightarrow\frac{7}{16}\rightarrow
\frac{24}{55}\rightarrow\frac{17}{39}\rightarrow
\frac{10}{23}\rightarrow\;\frac{23}{53}\rightarrow
\frac{13}{30}\rightarrow\frac{29}{67}\rightarrow
\frac{16}{37}\rightarrow\;\frac{3}{7}\rightarrow\;
\frac{8}{19}\rightarrow\frac{13}{31}\rightarrow
\frac{18}{43}\rightarrow\nonumber\\
&&\frac{5}{12}\rightarrow\frac{17}{41}\rightarrow
\frac{12}{29}\rightarrow\frac{7}{17}\rightarrow
\frac{16}{39}\rightarrow\;\frac{9}{22}\rightarrow
\frac{20}{49}\rightarrow\frac{11}{27}\rightarrow\;
\frac{2}{5}\rightarrow\;\frac{5}{13}\rightarrow
\frac{8}{21}\rightarrow\frac{11}{29}\rightarrow\;
\frac{3}{8}\rightarrow\nonumber\\
&&\frac{10}{27}\rightarrow\frac{7}{19}\rightarrow
\frac{4}{11}\rightarrow\frac{9}{25}\rightarrow
\frac{5}{14}\rightarrow\;\frac{11}{31}\rightarrow
\frac{6}{17}\rightarrow\;\frac{1}{3}\rightarrow\;
\frac{6}{19}\rightarrow\frac{11}{35}\rightarrow
\frac{5}{16}\rightarrow\frac{9}{29}\rightarrow
\frac{4}{13}\rightarrow\nonumber\\
&&\frac{7}{23}\rightarrow\frac{3}{10}\rightarrow
\frac{5}{17}\rightarrow\frac{2}{7}\rightarrow\;\;
\frac{3}{11}\rightarrow\;\frac{4}{15}\rightarrow
\frac{5}{19}\rightarrow\frac{6}{23}\rightarrow\;
\frac{1}{4}\rightarrow\;\frac{6}{25}\rightarrow
\frac{5}{21}\rightarrow\frac{4}{17}\rightarrow
\frac{3}{13}\rightarrow\nonumber\\
&&\frac{2}{9}\rightarrow\;\;\frac{5}{23}\rightarrow
\frac{3}{14}\rightarrow\frac{4}{19}\rightarrow
\frac{5}{24}\rightarrow\frac{6}{29}\rightarrow\;
\frac{1}{5}\rightarrow\;\frac{6}{31}\rightarrow\;
\frac{5}{26}\rightarrow\frac{4}{21}\rightarrow
\frac{3}{16}\rightarrow\frac{5}{27}\rightarrow
\frac{2}{11}\rightarrow\nonumber\\
&&\frac{3}{17}\rightarrow
\frac{4}{23}\rightarrow\;\frac{1}{6}\rightarrow
\frac{3}{19}\rightarrow\frac{2}{13}\rightarrow\;\;
\frac{1}{7}\rightarrow\;\frac{2}{15}\rightarrow\;
\frac{1}{8}\rightarrow\;\frac{2}{17}\rightarrow\;\;\frac{1}{9}.\nonumber 
\end{eqnarray}

\noindent Thus, we identify dual pairs of 
filling factors observed

\begin{eqnarray} 
\label{e.22}
(\nu,\tilde{\nu})&=&\left(\frac{1}{3},\frac{2}{3}\right), 
\left(\frac{5}{3},\frac{4}{3}\right), \left(\frac{1}{5},\frac{4}{5}\right), 
\left(\frac{2}{7},\frac{5}{7}\right),\left(\frac{2}{9},\frac{7}{9}\right), 
\left(\frac{2}{5},\frac{3}{5}\right),\left(\frac{7}{3},\frac{8}{3}\right),\nonumber\\
&&\left(\frac{3}{7},\frac{4}{7}\right), 
\left(\frac{4}{9},\frac{5}{9}\right), \left(\frac{8}{5},\frac{7}{5}\right),
\left(\frac{6}{13},\frac{7}{13}\right),\left(\frac{5}{11},\frac{6}{11}\right),
\left(\frac{10}{7},\frac{11}{7}\right),\\
&&\left(\frac{4}{11},\frac{7}{11}\right),
\left(\frac{13}{4},\frac{15}{4}\right), \left(\frac{14}{5},\frac{11}{5}\right),
\left(\frac{9}{4},\frac{11}{4}\right),\left(\frac{19}{7},\frac{16}{7}\right) 
etc.\nonumber
\end{eqnarray}

\noindent and other ones to be verified as 

\begin{eqnarray} 
(\nu,{\tilde{\nu}})&=&\left(\frac{2}{11},\frac{9}{11}\right),
\left(\frac{3}{19},\frac{16}{19}\right), 
\left(\frac{3}{17},\frac{14}{17}\right), \left(\frac{2}{13},\frac{11}{13}\right), 
\left(\frac{1}{7},\frac{6}{7}\right), \left(\frac{2}{15},\frac{13}{15}\right),\\
&&\left(\frac{5}{13},\frac{8}{13}\right),
\left(\frac{4}{13},\frac{9}{13}\right),\left(\frac{5}{17},\frac{12}{17}\right), 
\left(\frac{26}{7},\frac{23}{7}\right),\left(\frac{19}{5},\frac{16}{5}\right), 
\left(\frac{34}{9},\frac{29}{9}\right) etc.\nonumber
\end{eqnarray}

Here, we observe that our approach, in terms of fractal sets of 
dual topological filling factors, embodies 
the structure of the modular group as discussed in the literature
\cite{R7,R28} and the quantum Hall transitions satisfy some 
properties related with the Farey sequences of rational numbers. The 
transitions allowed are those generated by the condition
 $\mid p_{2}q_{1}
-p_{1}q_{2}\mid=1$, 
with $\nu_{1}=\frac{p_{1}}{q_{1}}$ and $\nu_{2}=
\frac{p_{2}}{q_{2}}$. In\cite{R28} the properties of integer and fractional 
quantum Hall effect were considered in terms of a subgroup of the modular group 
$SL(2,{\bf Z})$, such that the group acts on the upper-half complex plane 
parameterised by a complex conductivity 
and so generates the phase diagram of the quantum Hall effect. The rules 
obtained there are the same that our formulation\cite{R7} has extracted with the bonus 
that we define the universality classes of the quantum Hall 
transitions in terms of fractal sets labelled by a fractal parameter, 
besides all our theoretical description which consider the Hall 
states modelled by systems of fractons with a specific value of spin.

We verify distinct 
possible transitions for the FQHE and we define families of 
universality classes to them, for example, we have the groups:\\
\\

{\bf Group I}

\begin{eqnarray}
&&\biggl\{\frac{1}{5},\frac{9}{5},\frac{11}{5},\frac{19}{5},
\cdots\biggr\}_{h=\frac{9}{5}}\rightarrow\;\;
\biggl\{\frac{2}{9},\frac{16}{9},\frac{20}{9},\frac{34}{9},
\cdots\biggr\}_{h=\frac{16}{9}}\rightarrow
\biggl\{\frac{1}{4},\frac{7}{4},\frac{9}{4},\frac{15}{4},
\cdots\biggr\}_{h=\frac{7}{4}}\rightarrow\nonumber\\ 
&&\biggl\{\frac{2}{7},\frac{12}{7},\frac{16}{7},\frac{26}{7},
\cdots\biggr\}_{h=\frac{12}{7}}\rightarrow 
\biggl\{\frac{1}{3},\frac{5}{3},\frac{7}{3},\frac{11}{3},
\cdots\biggr\}_{h=\frac{5}{3}}\rightarrow\;\;\;\;
\biggl\{\frac{2}{5},\frac{8}{5},\frac{12}{5},\frac{18}{5},
\cdots\biggr\}_{h=\frac{8}{5}}\rightarrow\nonumber\\
&&\biggl\{\frac{3}{7},\frac{11}{7},\frac{17}{7},\frac{25}{7},
\cdots\biggr\}_{h=\frac{11}{7}}
\rightarrow\biggl\{\frac{1}{2},\frac{3}{2},
\frac{5}{2},\frac{7}{2},\cdots\biggr\}_{h=\frac{3}{2}}\;\;\;\;\;\;
\rightarrow\biggl\{\frac{4}{7},\frac{10}{7},\frac{18}{7},\frac{24}{7},
\cdots\biggr\}_{h=\frac{10}{7}}\rightarrow\\
&&\biggl\{\frac{3}{5},\frac{7}{5},\frac{13}{5},\frac{17}{5},
\cdots\biggr\}_{h=\frac{7}{5}}\rightarrow\;\;\;
\biggl\{\frac{2}{3},\frac{4}{3},\frac{8}{3},\frac{10}{3},
\cdots\biggr\}_{h=\frac{4}{3}}
\rightarrow\biggl\{\frac{5}{7},\frac{9}{7},\frac{21}{7},\frac{23}{7},
\cdots\biggr\}_{h=\frac{9}{7}}
\rightarrow\nonumber\\
&&\biggl\{\frac{3}{4},\frac{5}{4},\frac{11}{4},\frac{13}{4},
\cdots\biggr\}_{h=\frac{5}{4}}\rightarrow\;\;\;
\biggl\{\frac{7}{9},\frac{11}{9},\frac{25}{9},
\frac{29}{9},\cdots\biggr\}_{h=\frac{11}{9}}
\rightarrow
\biggl\{\frac{4}{5},\frac{6}{5},\frac{14}{5},\frac{16}{5},
\cdots\biggr\}_{h=\frac{6}{5}}.\nonumber
\end{eqnarray}

\noindent {\bf Group II}

\begin{eqnarray}
&&\biggl\{\frac{4}{5},\frac{6}{5},\frac{14}{5},\frac{16}{5},
\cdots\biggr\}_{h=\frac{6}{5}}\rightarrow\;\;\;\;
\biggl\{\frac{7}{9},\frac{11}{9},\frac{25}{9},\frac{29}{9},
\cdots\biggr\}_{h=\frac{11}{9}}\rightarrow\;\;
 \biggl\{\frac{3}{4},\frac{5}{4},\frac{11}{4},\frac{13}{4},
\cdots\biggr\}_{h=\frac{5}{4}}\rightarrow\nonumber\\
&&\biggl\{\frac{5}{7},\frac{9}{7},\frac{19}{7},\frac{23}{7},
\cdots\biggr\}_{h=\frac{9}{7}}
\rightarrow\;\;\;\;
\biggl\{\frac{2}{3},\frac{4}{3},\frac{8}{3},\frac{10}{3},
\cdots\biggr\}_{h=\frac{4}{3}}
\rightarrow\;\;\;\;\;\;\; 
\biggl\{\frac{3}{5},\frac{7}{5},\frac{13}{5},\frac{17}{5},
\cdots\biggr\}_{h=\frac{7}{5}}
\rightarrow\nonumber\\
&&\biggl\{\frac{4}{7},\frac{10}{7},\frac{18}{7},\frac{24}{7},
\cdots\biggr\}_{h=\frac{10}{7}}
\rightarrow\;\;
\biggl\{\frac{5}{9},\frac{13}{9},\frac{23}{9},\frac{31}{9},
\cdots\biggr\}_{h=\frac{13}{9}}\rightarrow\;  
\biggl\{\frac{6}{11},\frac{16}{11},\frac{28}{11},\frac{38}{11},
\cdots\biggr\}_{h=\frac{16}{11}}\rightarrow\nonumber\\
&&\biggl\{\frac{7}{13},\frac{19}{13},\frac{33}{13},\frac{45}{13},
\cdots\biggr\}_{h=\frac{19}{13}}
\rightarrow
\biggl\{\frac{1}{2},\frac{3}{2},\frac{5}{2},\frac{7}{2},
\cdots\biggr\}_{h=\frac{3}{2}}
\rightarrow\;\;\;\;\;\;\;\; 
\biggl\{\frac{6}{13},\frac{20}{13},\frac{32}{13},\frac{46}{13},
\cdots\biggr\}_{h=\frac{20}{13}}
\rightarrow\\
&&\biggl\{\frac{5}{11},\frac{17}{11},\frac{27}{11},\frac{39}{11},
\cdots\biggr\}_{h=\frac{17}{11}}
\rightarrow
\biggl\{\frac{4}{9},\frac{14}{9},\frac{22}{9},\frac{32}{9},
\cdots\biggr\}_{h=\frac{14}{9}}\rightarrow\;\;  
\biggl\{\frac{3}{7},\frac{11}{7},\frac{17}{7},\frac{25}{7},
\cdots\biggr\}_{h=\frac{11}{7}}\rightarrow\nonumber\\
&&\biggl\{\frac{2}{5},\frac{8}{5},\frac{12}{5},\frac{18}{5},
\cdots\biggr\}_{h=\frac{8}{5}}
\rightarrow\;\;\;\;\;
\biggl\{\frac{1}{3},\frac{5}{3},\frac{7}{3},\frac{11}{3},
\cdots\biggr\}_{h=\frac{5}{3}}
\rightarrow\;\;\;\;\;\; 
\biggl\{\frac{2}{7},\frac{12}{7},\frac{16}{7},\frac{26}{7},
\cdots\biggr\}_{h=\frac{12}{7}}
\rightarrow\nonumber\\
&&\biggl\{\frac{1}{4},\frac{7}{4},\frac{9}{4},\frac{15}{4},
\cdots\biggr\}_{h=\frac{7}{4}}\rightarrow\;\;\;\;\;\;
\biggl\{\frac{2}{9},\frac{16}{9},\frac{20}{9},\frac{34}{9},
\cdots\biggr\}_{h=\frac{16}{9}}\rightarrow\;\;
\biggl\{\frac{1}{5},\frac{9}{5},\frac{11}{5},\frac{19}{5},
\cdots\biggr\}_{h=\frac{9}{5}}
.\nonumber
\end{eqnarray}

\newpage

\noindent {\bf Group III}

\begin{eqnarray}
&&\biggl\{\frac{1}{8},\frac{15}{8},\frac{17}{8},\frac{31}{8},
\cdots\biggr\}_{h=\frac{15}{8}}\rightarrow\;\;\;
\biggl\{\frac{3}{23},\frac{43}{23},\frac{49}{23},\frac{89}{23},
\cdots\biggr\}_{h=\frac{43}{23}}\rightarrow\;
\biggl\{\frac{2}{15},\frac{28}{15},\frac{32}{15},\frac{58}{15},
\cdots\biggr\}_{h=\frac{28}{15}}\rightarrow\nonumber\\
&&\biggl\{\frac{1}{7},\frac{13}{7},\frac{15}{7},\frac{27}{7},
\cdots\biggr\}_{h=\frac{13}{7}}
\rightarrow\;\;\;
\biggl\{\frac{1}{6},\frac{11}{6},\frac{13}{6},\frac{23}{6},
\cdots\biggr\}_{h=\frac{11}{6}}
\rightarrow\;\;\;
\biggl\{\frac{1}{5},\frac{9}{5},\frac{11}{5},\frac{19}{5},
\cdots\biggr\}_{h=\frac{9}{5}}
\rightarrow\rightarrow\nonumber\\
&&\biggl\{\frac{2}{9},\frac{16}{9},\frac{20}{9},\frac{34}{9},
\cdots\biggr\}_{h=\frac{16}{9}}
\rightarrow\;\;\;
\biggl\{\frac{1}{4},\frac{7}{4},\frac{9}{4},\frac{15}{4},
\cdots\biggr\}_{h=\frac{7}{4}}\rightarrow\;\;\;\;\;\;\;
\biggl\{\frac{3}{11},\frac{19}{11},\frac{25}{11},\frac{41}{11},
\cdots\biggr\}_{h=\frac{19}{11}}\rightarrow\;\nonumber\\
&&\biggl\{\frac{2}{7},\frac{12}{7},\frac{16}{7},\frac{26}{7},
\cdots\biggr\}_{h=\frac{12}{7}}
\rightarrow\;\;\;
\biggl\{\frac{3}{10},\frac{17}{10},\frac{23}{10},\frac{37}{10},
\cdots\biggr\}_{h=\frac{17}{10}}
\rightarrow\;
\biggl\{\frac{1}{3},\frac{5}{3},\frac{7}{3},\frac{11}{3},
\cdots\biggr\}_{h=\frac{5}{3}}
\rightarrow\nonumber\\
&&\biggl\{\frac{2}{5},\frac{8}{5},\frac{12}{5},\frac{18}{5},
\cdots\biggr\}_{h=\frac{8}{5}}\rightarrow\;\;\;\;\;\;
\biggl\{\frac{3}{7},\frac{11}{7},\frac{17}{7},\frac{25}{7},
\cdots\biggr\}_{h=\frac{11}{7}}
\rightarrow\;\;\;
\biggl\{\frac{4}{9},\frac{14}{9},\frac{22}{9},\frac{32}{9},
\cdots\biggr\}_{h=\frac{14}{9}}
\rightarrow\nonumber\\
&&\biggl\{\frac{5}{11},\frac{17}{11},\frac{27}{11},\frac{39}{11},
\cdots\biggr\}_{h=\frac{17}{11}}\rightarrow\;
\biggl\{\frac{6}{13},\frac{20}{13},\frac{32}{13},\frac{46}{13},
\cdots\biggr\}_{h=\frac{20}{13}}
\rightarrow\;\;
\biggl\{\frac{1}{2},\frac{3}{2},\frac{5}{2},\frac{7}{2},
\cdots\biggr\}_{h=\frac{3}{2}}
\rightarrow\nonumber\\
&&\biggl\{\frac{7}{13},\frac{19}{13},\frac{33}{13},\frac{45}{13},
\cdots\biggr\}_{h=\frac{19}{13}}\rightarrow\;
\biggl\{\frac{6}{11},\frac{16}{11},\frac{28}{11},\frac{38}{11},
\cdots\biggr\}_{h=\frac{16}{11}}
\rightarrow\;\;
\biggl\{\frac{5}{9},\frac{13}{9},\frac{23}{9},\frac{31}{9},
\cdots\biggr\}_{h=\frac{13}{9}}
\rightarrow\\
&&\biggl\{\frac{4}{7},\frac{10}{7},\frac{18}{7},\frac{24}{7},
\cdots\biggr\}_{h=\frac{10}{7}}\rightarrow\;\;\;
\biggl\{\frac{3}{5},\frac{7}{5},\frac{13}{5},\frac{17}{5},
\cdots\biggr\}_{h=\frac{7}{5}}
\rightarrow\;\;\;\;\;\;
\biggl\{\frac{2}{3},\frac{4}{3},\frac{8}{3},\frac{10}{3},
\cdots\biggr\}_{h=\frac{4}{3}}
\rightarrow\nonumber\\
&&\biggl\{\frac{7}{10},\frac{13}{10},\frac{27}{10},\frac{33}{10},
\cdots\biggr\}_{h=\frac{13}{10}}\rightarrow\;
\biggl\{\frac{5}{7},\frac{9}{7},\frac{19}{7},\frac{23}{7},
\cdots\biggr\}_{h=\frac{9}{7}}
\rightarrow\;\;\;\;\;\;
\biggl\{\frac{8}{11},\frac{14}{11},\frac{30}{11},\frac{36}{11},
\cdots\biggr\}_{h=\frac{14}{11}}
\rightarrow\nonumber\\
&&\biggl\{\frac{3}{4},\frac{5}{4},\frac{11}{4},\frac{13}{4},
\cdots\biggr\}_{h=\frac{5}{4}}\rightarrow\;\;\;\;\;\;
\biggl\{\frac{7}{9},\frac{11}{9},\frac{25}{9},\frac{29}{9},
\cdots\biggr\}_{h=\frac{11}{9}}
\rightarrow\;\;\;
\biggl\{\frac{4}{5},\frac{6}{5},\frac{14}{5},\frac{16}{5},
\cdots\biggr\}_{h=\frac{6}{5}}
\rightarrow\nonumber\\
&&\biggl\{\frac{5}{6},\frac{7}{6},\frac{17}{6},\frac{19}{6},
\cdots\biggr\}_{h=\frac{7}{6}}\rightarrow\;\;\;\;\;\;
\biggl\{\frac{6}{7},\frac{8}{7},\frac{20}{7},\frac{22}{7},
\cdots\biggr\}_{h=\frac{8}{7}}
\rightarrow\;\;\;\;\;\;
\biggl\{\frac{13}{15},\frac{17}{15},\frac{43}{15},\frac{47}{15},
\cdots\biggr\}_{h=\frac{17}{15}}
\rightarrow\nonumber\\
&&\biggl\{\frac{20}{23},\frac{26}{23},\frac{66}{23},\frac{72}{23},
\cdots\biggr\}_{h=\frac{26}{23}}\rightarrow\;
\biggl\{\frac{7}{8},\frac{9}{8},\frac{23}{8},\frac{25}{8},
\cdots\biggr\}_{h=\frac{9}{8}}.\nonumber
\end{eqnarray}

\newpage

\noindent{\bf Groupo IV}

\begin{eqnarray}
&&\biggl\{\frac{8}{9},\frac{10}{9},\frac{26}{9},\frac{28}{9},
\cdots\biggr\}_{h=\frac{10}{9}}
\rightarrow\;\;\;
\biggl\{\frac{15}{17},\frac{19}{17},\frac{49}{17},\frac{53}{17},
\cdots\biggr\}_{h=\frac{19}{17}}
\rightarrow\biggl\{\frac{7}{8},\frac{9}{8},\frac{23}{8},\frac{25}{8},
\cdots\biggr\}_{h=\frac{9}{8}}\rightarrow\nonumber\\
&&\biggl\{\frac{13}{15},\frac{17}{15},\frac{43}{15},\frac{47}{15},
\cdots\biggr\}_{h=\frac{17}{15}}
\rightarrow\;
\biggl\{\frac{6}{7},\frac{8}{7},\frac{20}{7},\frac{22}{7},
\cdots\biggr\}_{h=\frac{8}{7}}
\rightarrow\;\;\;\;\;\biggl\{\frac{11}{13},\frac{15}{13},\frac{37}{13},\frac{41}{13},
\cdots\biggr\}_{h=\frac{15}{13}}\rightarrow\nonumber\\
&&\biggl\{\frac{16}{19},\frac{22}{19},\frac{54}{19},\frac{60}{19},
\cdots\biggr\}_{h=\frac{22}{19}}
\rightarrow\;
\biggl\{\frac{5}{6},\frac{7}{6},\frac{17}{6},\frac{19}{6},
\cdots\biggr\}_{h=\frac{7}{6}}
\rightarrow\;\;\;\;\;\biggl\{\frac{19}{23},\frac{27}{23},\frac{65}{23},\frac{73}{23},
\cdots\biggr\}_{h=\frac{27}{23}}\rightarrow\nonumber\\
&&\biggl\{\frac{14}{17},\frac{20}{17},\frac{48}{17},\frac{54}{17},
\cdots\biggr\}_{h=\frac{20}{17}}
\rightarrow\;
\biggl\{\frac{9}{11},\frac{13}{11},\frac{31}{11},\frac{35}{11},
\cdots\biggr\}_{h=\frac{13}{11}}
\rightarrow
\biggl\{\frac{22}{27},\frac{32}{27},\frac{76}{27},\frac{86}{27},
\cdots\biggr\}_{h=\frac{32}{27}}
\rightarrow\nonumber\\
&&\biggl\{\frac{13}{16},\frac{19}{16},\frac{45}{16},\frac{51}{16},
\cdots\biggr\}_{h=\frac{19}{16}}
\rightarrow\;
\biggl\{\frac{17}{21},\frac{25}{21},\frac{59}{21},\frac{67}{21},
\cdots\biggr\}_{h=\frac{25}{21}}
\rightarrow
\biggl\{\frac{21}{26},\frac{31}{26},\frac{73}{26},\frac{83}{26},
\cdots\biggr\}_{h=\frac{31}{26}}
\rightarrow\nonumber\\
&&\biggl\{\frac{25}{31},\frac{37}{31},\frac{87}{31},\frac{99}{31},
\cdots\biggr\}_{h=\frac{37}{31}}
\rightarrow\;
\biggl\{\frac{4}{5},\frac{6}{5},\frac{14}{5},\frac{16}{5},
\cdots\biggr\}_{h=\frac{6}{5}}\rightarrow\;\;\;\;\;
\biggl\{\frac{23}{29},\frac{35}{29},\frac{81}{29},\frac{93}{29},
\cdots\biggr\}_{h=\frac{35}{29}}
\rightarrow\nonumber\\
&&\biggl\{\frac{19}{24},\frac{29}{24},\frac{67}{24},\frac{77}{24},
\cdots\biggr\}_{h=\frac{29}{24}}
\rightarrow\;
\biggl\{\frac{15}{19},\frac{23}{19},\frac{53}{19},\frac{61}{19},
\cdots\biggr\}_{h=\frac{23}{19}}\rightarrow
\biggl\{\frac{11}{14},\frac{17}{14},\frac{39}{14},\frac{45}{14},
\cdots\biggr\}_{h=\frac{17}{14}}
\rightarrow\nonumber\\
&&\biggl\{\frac{18}{23},\frac{28}{23},\frac{64}{23},\frac{74}{23},
\cdots\biggr\}_{h=\frac{28}{23}}
\rightarrow\;
\biggl\{\frac{7}{9},\frac{11}{9},\frac{25}{9},\frac{29}{9},
\cdots\biggr\}_{h=\frac{11}{9}}
\rightarrow\;\;
\biggl\{\frac{10}{13},\frac{16}{13},\frac{36}{13},\frac{42}{13},
\cdots\biggr\}_{h=\frac{16}{13}}
\rightarrow\nonumber\\
&&\biggl\{\frac{13}{17},\frac{21}{17},\frac{47}{17},\frac{55}{17},
\cdots\biggr\}_{h=\frac{21}{17}}\rightarrow\;
\biggl\{\frac{16}{21},\frac{26}{21},\frac{58}{21},\frac{68}{21},
\cdots\biggr\}_{h=\frac{26}{21}}
\rightarrow
\biggl\{\frac{19}{25},\frac{31}{25},\frac{69}{25},\frac{81}{25},
\cdots\biggr\}_{h=\frac{31}{25}}
\rightarrow\nonumber\\
&&\biggl\{\frac{3}{4},\frac{5}{4},\frac{11}{4},\frac{13}{4},
\cdots\biggr\}_{h=\frac{5}{4}}\rightarrow\;\;\;\;\;\;
\biggl\{\frac{17}{23},\frac{29}{23},\frac{63}{23},\frac{75}{23},
\cdots\biggr\}_{h=\frac{29}{23}}\rightarrow
\biggl\{\frac{14}{19},\frac{24}{19},\frac{52}{19},\frac{62}{19},
\cdots\biggr\}_{h=\frac{24}{19}}
\rightarrow\nonumber\\
&&\biggl\{\frac{11}{15},\frac{19}{15},\frac{41}{15},\frac{49}{15},
\cdots\biggr\}_{h=\frac{19}{15}}
\rightarrow\;
\biggl\{\frac{8}{11},\frac{14}{11},\frac{30}{11},\frac{36}{11},
\cdots\biggr\}_{h=\frac{14}{11}}\rightarrow
\biggl\{\frac{5}{7},\frac{9}{7},\frac{19}{7},\frac{23}{7},
\cdots\biggr\}_{h=\frac{9}{7}}\rightarrow\nonumber\\
&&\biggl\{\frac{12}{17},\frac{22}{17},\frac{46}{17},\frac{56}{17},
\cdots\biggr\}_{h=\frac{22}{17}}\rightarrow\;
\biggl\{\frac{7}{10},\frac{13}{10},\frac{27}{10},\frac{33}{10},
\cdots\biggr\}_{h=\frac{13}{10}}
\rightarrow
\biggl\{\frac{16}{23},\frac{30}{23},\frac{62}{23},\frac{76}{23},
\cdots\biggr\}_{h=\frac{30}{23}}
\rightarrow\nonumber\\
&&\biggl\{\frac{9}{13},\frac{17}{13},\frac{35}{13},\frac{43}{13},
\cdots\biggr\}_{h=\frac{17}{13}}
\rightarrow\;
\biggl\{\frac{20}{29},\frac{38}{29},\frac{78}{29},\frac{96}{29},
\cdots\biggr\}_{h=\frac{38}{29}}
\rightarrow
\biggl\{\frac{11}{16},\frac{21}{16},\frac{43}{16},\frac{53}{16},
\cdots\biggr\}_{h=\frac{21}{16}}
\rightarrow\nonumber\\
&&\biggl\{\frac{24}{35},\frac{46}{35},\frac{94}{35},\frac{116}{35},
\cdots\biggr\}_{h=\frac{46}{35}}
\rightarrow
\biggl\{\frac{13}{19},\frac{25}{19},\frac{51}{19},\frac{63}{19},
\cdots\biggr\}_{h=\frac{25}{19}}
\rightarrow
\biggl\{\frac{2}{3},\frac{4}{3},\frac{8}{3},\frac{10}{3},
\cdots\biggr\}_{h=\frac{4}{3}}
\rightarrow\nonumber\\
&&\biggl\{\frac{11}{17},\frac{23}{17},\frac{45}{17},\frac{57}{17},
\cdots\biggr\}_{h=\frac{23}{17}}
\rightarrow\;
\biggl\{\frac{20}{31},\frac{42}{31},\frac{82}{31},\frac{104}{31},
\cdots\biggr\}_{h=\frac{42}{31}}
\rightarrow
\biggl\{\frac{9}{14},\frac{19}{14},\frac{37}{14},\frac{47}{14},
\cdots\biggr\}_{h=\frac{19}{14}}
\rightarrow\nonumber\\
&&\biggl\{\frac{16}{25},\frac{34}{25},\frac{66}{25},\frac{84}{25},
\cdots\biggr\}_{h=\frac{34}{25}}
\rightarrow\;
\biggl\{\frac{7}{11},\frac{15}{11},\frac{29}{11},\frac{37}{11},
\cdots\biggr\}_{h=\frac{15}{11}}
\rightarrow\;\;
\biggl\{\frac{12}{19},\frac{26}{19},\frac{50}{19},\frac{64}{19},
\cdots\biggr\}_{h=\frac{26}{19}}
\rightarrow\nonumber\\
&&\biggl\{\frac{17}{27},\frac{37}{27},\frac{71}{27},\frac{91}{27},
\cdots\biggr\}_{h=\frac{37}{27}}
\rightarrow\;
\biggl\{\frac{5}{8},\frac{11}{8},\frac{21}{8},\frac{27}{8},
\cdots\biggr\}_{h=\frac{11}{8}}
\rightarrow\;\;\;
\biggl\{\frac{18}{29},\frac{40}{29},\frac{76}{29},\frac{98}{29},
\cdots\biggr\}_{h=\frac{40}{29}}
\rightarrow\nonumber\\
&&\biggl\{\frac{13}{21},\frac{29}{21},\frac{55}{21},\frac{71}{21},
\cdots\biggr\}_{h=\frac{29}{21}}
\rightarrow\;
\biggl\{\frac{8}{13},\frac{18}{13},\frac{34}{13},\frac{44}{13},
\cdots\biggr\}_{h=\frac{18}{13}}
\rightarrow\;
\biggl\{\frac{3}{5},\frac{7}{5},\frac{13}{5},\frac{17}{5},
\cdots\biggr\}_{h=\frac{7}{5}}
\rightarrow\nonumber\\
&&\biggl\{\frac{16}{27},\frac{38}{27},\frac{70}{27},\frac{92}{27},
\cdots\biggr\}_{h=\frac{38}{27}}
\rightarrow
\biggl\{\frac{29}{49},\frac{69}{49},\frac{127}{49},\frac{167}{49},
\cdots\biggr\}_{h=\frac{69}{49}}
\rightarrow
\biggl\{\frac{13}{22},\frac{31}{22},\frac{57}{22},\frac{75}{22},
\cdots\biggr\}_{h=\frac{31}{22}}
\rightarrow\nonumber\\
&&\biggl\{\frac{23}{39},\frac{55}{39},\frac{81}{39},\frac{113}{39},
\cdots\biggr\}_{h=\frac{55}{39}}
\rightarrow\;\;
\biggl\{\frac{10}{17},\frac{24}{17},\frac{44}{17},\frac{58}{17},
\cdots\biggr\}_{h=\frac{24}{17}}
\rightarrow\;\;\;
\biggl\{\frac{17}{29},\frac{41}{29},\frac{75}{29},\frac{99}{29},
\cdots\biggr\}_{h=\frac{41}{29}}
\rightarrow\nonumber\\
&&\biggl\{\frac{24}{41},\frac{58}{41},\frac{106}{41},\frac{140}{41},
\cdots\biggr\}_{h=\frac{58}{41}}
\rightarrow
\biggl\{\frac{7}{12},\frac{17}{12},\frac{31}{12},\frac{41}{12},
\cdots\biggr\}_{h=\frac{17}{12}}
\rightarrow\;\;\;
\biggl\{\frac{25}{43},\frac{61}{43},\frac{111}{43},\frac{147}{43},
\cdots\biggr\}_{h=\frac{61}{43}}
\rightarrow\nonumber\\
&&\biggl\{\frac{18}{31},\frac{44}{31},\frac{80}{31},\frac{106}{31},
\cdots\biggr\}_{h=\frac{44}{31}}
\rightarrow\;\;
\biggl\{\frac{11}{19},\frac{27}{19},\frac{49}{19},\frac{65}{19},
\cdots\biggr\}_{h=\frac{27}{19}}
\rightarrow\;\;\;
\biggl\{\frac{4}{7},\frac{10}{7},\frac{18}{7},\frac{24}{7},
\cdots\biggr\}_{h=\frac{10}{7}}
\rightarrow\nonumber\\
&&\biggl\{\frac{21}{37},\frac{53}{37},\frac{95}{37},\frac{127}{37},
\cdots\biggr\}_{h=\frac{53}{37}}
\rightarrow\;\;
\biggl\{\frac{38}{67},\frac{96}{67},\frac{172}{67},\frac{230}{67},
\cdots\biggr\}_{h=\frac{96}{67}}
\rightarrow
\biggl\{\frac{17}{30},\frac{43}{30},\frac{77}{30},\frac{103}{30},
\cdots\biggr\}_{h=\frac{43}{30}}
\rightarrow\nonumber\\
&&\biggl\{\frac{30}{53},\frac{76}{53},\frac{136}{53},\frac{182}{53},
\cdots\biggr\}_{h=\frac{76}{53}}
\rightarrow
\biggl\{\frac{13}{23},\frac{33}{23},\frac{59}{23},\frac{79}{23},
\cdots\biggr\}_{h=\frac{33}{23}}
\rightarrow\;\;\;\;
\biggl\{\frac{32}{39},\frac{56}{39},\frac{100}{39},\frac{134}{39},
\cdots\biggr\}_{h=\frac{56}{39}}
\rightarrow\nonumber\\
&&\biggl\{\frac{31}{55},\frac{79}{55},\frac{141}{55},\frac{189}{55},
\cdots\biggr\}_{h=\frac{79}{55}}
\rightarrow
\biggl\{\frac{9}{16},\frac{23}{16},\frac{41}{16},\frac{65}{16},
\cdots\biggr\}_{h=\frac{23}{16}}
\rightarrow\;\;\;\;
\biggl\{\frac{32}{57},\frac{82}{57},\frac{146}{57},\frac{196}{57},
\cdots\biggr\}_{h=\frac{82}{57}}
\rightarrow\nonumber\\
&&\biggl\{\frac{23}{41},\frac{59}{41},\frac{105}{41},\frac{141}{41},
\cdots\biggr\}_{h=\frac{59}{41}}
\rightarrow
\biggl\{\frac{14}{25},\frac{36}{25},\frac{64}{25},\frac{86}{25},
\cdots\biggr\}_{h=\frac{36}{25}}
\rightarrow\;\;\;\;
\biggl\{\frac{5}{9},\frac{13}{9},\frac{23}{9},\frac{31}{9},
\cdots\biggr\}_{h=\frac{13}{9}}
\rightarrow\nonumber\\
&&\biggl\{\frac{6}{11},\frac{16}{11},\frac{28}{11},\frac{38}{11},
\cdots\biggr\}_{h=\frac{16}{11}}
\rightarrow\;\;\;
\biggl\{\frac{7}{13},\frac{19}{13},\frac{33}{13},\frac{45}{13},
\cdots\biggr\}_{h=\frac{19}{13}}
\rightarrow\;\;\;\;\;
\biggl\{\frac{8}{15},\frac{22}{15},\frac{38}{15},\frac{52}{15},
\cdots\biggr\}_{h=\frac{22}{15}}
\rightarrow\nonumber\\
&&\biggl\{\frac{9}{17},\frac{25}{17},\frac{43}{17},\frac{59}{17},
\cdots\biggr\}_{h=\frac{25}{17}}
\rightarrow\;\;\;
\biggl\{\frac{10}{19},\frac{28}{19},\frac{48}{19},\frac{66}{19},
\cdots\biggr\}_{h=\frac{28}{19}}
\rightarrow\;\;\;\;\;
\biggl\{\frac{11}{21},\frac{31}{21},\frac{53}{21},\frac{73}{21},
\cdots\biggr\}_{h=\frac{31}{21}}
\rightarrow\nonumber\\
&&\biggl\{\frac{1}{2},\frac{3}{2},\frac{5}{2},\frac{7}{2},
\cdots\biggr\}_{h=\frac{3}{2}}
\rightarrow\;\;\;\;\;\;\;\;\;\;\;
\biggl\{\frac{10}{21},\frac{32}{21},\frac{52}{21},\frac{74}{21},
\cdots\biggr\}_{h=\frac{32}{21}}
\rightarrow\;\;\;\;\;
\biggl\{\frac{9}{19},\frac{29}{19},\frac{47}{19},\frac{67}{19},
\cdots\biggr\}_{h=\frac{29}{19}}
\rightarrow\nonumber\\
&&\biggl\{\frac{8}{17},\frac{26}{17},\frac{42}{17},\frac{60}{17},
\cdots\biggr\}_{h=\frac{26}{17}}
\rightarrow\;\;\;
\biggl\{\frac{7}{15},\frac{23}{15},\frac{37}{15},\frac{53}{15},
\cdots\biggr\}_{h=\frac{23}{15}}
\rightarrow\;\;\;\;\;
\biggl\{\frac{6}{13},\frac{20}{13},\frac{32}{13},\frac{46}{13},
\cdots\biggr\}_{h=\frac{20}{13}}
\rightarrow\nonumber\\
&&\biggl\{\frac{5}{11},\frac{17}{11},\frac{27}{11},\frac{39}{11},
\cdots\biggr\}_{h=\frac{17}{11}}
\rightarrow\;\;\;
\biggl\{\frac{4}{9},\frac{14}{9},\frac{22}{9},\frac{32}{9},
\cdots\biggr\}_{h=\frac{14}{9}}
\rightarrow\;\;\;\;\;\;\;
\biggl\{\frac{11}{25},\frac{39}{25},\frac{61}{25},\frac{89}{25},
\cdots\biggr\}_{h=\frac{39}{25}}
\rightarrow\nonumber\\
&&\biggl\{\frac{18}{41},\frac{64}{41},\frac{100}{41},\frac{146}{41},
\cdots\biggr\}_{h=\frac{64}{41}}
\rightarrow
\biggl\{\frac{25}{57},\frac{89}{57},\frac{139}{57},\frac{203}{57},
\cdots\biggr\}_{h=\frac{89}{57}}
\rightarrow\;
\biggl\{\frac{7}{16},\frac{25}{16},\frac{39}{16},\frac{57}{16},
\cdots\biggr\}_{h=\frac{25}{16}}
\rightarrow\nonumber\\
&&\biggl\{\frac{24}{55},\frac{86}{55},\frac{134}{55},\frac{196}{55},
\cdots\biggr\}_{h=\frac{86}{55}}
\rightarrow
\biggl\{\frac{17}{39},\frac{61}{39},\frac{95}{39},\frac{139}{39},
\cdots\biggr\}_{h=\frac{61}{39}}
\rightarrow\;\;\;
\biggl\{\frac{10}{23},\frac{36}{23},\frac{56}{23},\frac{82}{23},
\cdots\biggr\}_{h=\frac{36}{23}}
\rightarrow\nonumber\\
&&\biggl\{\frac{23}{53},\frac{83}{53},\frac{129}{53},\frac{189}{53},
\cdots\biggr\}_{h=\frac{83}{53}}
\rightarrow
\biggl\{\frac{13}{30},\frac{47}{30},\frac{73}{30},\frac{107}{30},
\cdots\biggr\}_{h=\frac{47}{30}}
\rightarrow\;\;\;
\biggl\{\frac{29}{67},\frac{105}{67},\frac{163}{67},\frac{239}{67},
\cdots\biggr\}_{h=\frac{105}{67}}
\rightarrow\nonumber\\
&&\biggl\{\frac{16}{37},\frac{58}{37},\frac{90}{37},\frac{132}{37},
\cdots\biggr\}_{h=\frac{58}{37}}
\rightarrow\;\;
\biggl\{\frac{3}{7},\frac{11}{7},\frac{17}{7},\frac{25}{7},
\cdots\biggr\}_{h=\frac{11}{7}}
\rightarrow\;\;\;\;\;\;
\biggl\{\frac{8}{19},\frac{30}{19},\frac{46}{19},\frac{68}{19},
\cdots\biggr\}_{h=\frac{30}{19}}
\rightarrow\nonumber\\
&&\biggl\{\frac{13}{31},\frac{49}{31},\frac{75}{31},\frac{101}{31},
\cdots\biggr\}_{h=\frac{49}{31}}
\rightarrow\;\;
\biggl\{\frac{18}{43},\frac{68}{43},\frac{104}{43},\frac{154}{43},
\cdots\biggr\}_{h=\frac{68}{43}}
\rightarrow\;
\biggl\{\frac{5}{12},\frac{19}{12},\frac{29}{12},\frac{43}{12},
\cdots\biggr\}_{h=\frac{19}{12}}
\rightarrow\nonumber\\
&&\biggl\{\frac{17}{41},\frac{65}{41},\frac{99}{41},\frac{147}{41},
\cdots\biggr\}_{h=\frac{65}{41}}
\rightarrow\;\;
\biggl\{\frac{12}{29},\frac{46}{29},\frac{70}{29},\frac{104}{29},
\cdots\biggr\}_{h=\frac{46}{29}}
\rightarrow\;\;\;
\biggl\{\frac{7}{17},\frac{27}{17},\frac{41}{17},\frac{61}{17},
\cdots\biggr\}_{h=\frac{27}{17}}
\rightarrow\nonumber\\
&&\biggl\{\frac{16}{39},\frac{62}{39},\frac{94}{39},\frac{140}{39},
\cdots\biggr\}_{h=\frac{62}{39}}
\rightarrow\;\;
\biggl\{\frac{9}{22},\frac{35}{22},\frac{53}{22},\frac{79}{22},
\cdots\biggr\}_{h=\frac{35}{22}}
\rightarrow\;\;\;\;\;
\biggl\{\frac{20}{49},\frac{78}{49},\frac{118}{49},\frac{176}{49},
\cdots\biggr\}_{h=\frac{78}{49}}
\rightarrow\nonumber\\
&&\biggl\{\frac{11}{27},\frac{43}{27},\frac{65}{27},\frac{97}{27},
\cdots\biggr\}_{h=\frac{43}{27}}
\rightarrow\;\;\;\;
\biggl\{\frac{2}{5},\frac{8}{5},\frac{12}{5},\frac{18}{5},
\cdots\biggr\}_{h=\frac{8}{5}}
\rightarrow\;\;\;\;\;\;\;\;\;
\biggl\{\frac{5}{13},\frac{21}{13},\frac{31}{13},\frac{47}{13},
\cdots\biggr\}_{h=\frac{21}{13}}
\rightarrow\nonumber\\
&&\biggl\{\frac{8}{21},\frac{34}{21},\frac{50}{21},\frac{76}{21},
\cdots\biggr\}_{h=\frac{34}{21}}
\rightarrow\;\;
\biggl\{\frac{11}{29},\frac{47}{29},\frac{69}{29},\frac{105}{29},
\cdots\biggr\}_{h=\frac{47}{29}}
\rightarrow\;\;
\biggl\{\frac{3}{8},\frac{13}{8},\frac{19}{8},\frac{29}{8},
\cdots\biggr\}_{h=\frac{13}{8}}
\rightarrow\nonumber\\
&&\biggl\{\frac{10}{27},\frac{44}{27},\frac{64}{27},\frac{98}{27},
\cdots\biggr\}_{h=\frac{44}{27}}
\rightarrow\;\;
\biggl\{\frac{7}{19},\frac{31}{19},\frac{45}{19},\frac{71}{19},
\cdots\biggr\}_{h=\frac{31}{19}}
\rightarrow\;\;\;
\biggl\{\frac{4}{11},\frac{18}{11},\frac{26}{11},\frac{40}{11},
\cdots\biggr\}_{h=\frac{18}{11}}
\rightarrow\nonumber\\
&&\biggl\{\frac{9}{25},\frac{41}{25},\frac{59}{25},\frac{91}{25},
\cdots\biggr\}_{h=\frac{41}{25}}
\rightarrow\;\;
\biggl\{\frac{5}{14},\frac{23}{14},\frac{33}{14},\frac{51}{14},
\cdots\biggr\}_{h=\frac{23}{14}}
\rightarrow\;\;\;
\biggl\{\frac{11}{31},\frac{51}{31},\frac{73}{31},\frac{113}{31},
\cdots\biggr\}_{h=\frac{51}{31}}
\rightarrow\nonumber\\
&&\biggl\{\frac{6}{17},\frac{28}{17},\frac{40}{17},\frac{62}{17},
\cdots\biggr\}_{h=\frac{28}{17}}
\rightarrow\;\;
\biggl\{\frac{1}{3},\frac{5}{3},\frac{7}{3},\frac{11}{3},
\cdots\biggr\}_{h=\frac{5}{3}}
\rightarrow\;\;\;\;\;\;\;\;\;
\biggl\{\frac{6}{19},\frac{32}{19},\frac{44}{19},\frac{70}{19},
\cdots\biggr\}_{h=\frac{32}{19}}
\rightarrow\nonumber\\
&&\biggl\{\frac{11}{35},\frac{59}{35},\frac{81}{35},\frac{129}{35},
\cdots\biggr\}_{h=\frac{59}{35}}
\rightarrow
\biggl\{\frac{5}{16},\frac{27}{16},\frac{37}{16},\frac{59}{16},
\cdots\biggr\}_{h=\frac{27}{16}}
\rightarrow\;\;\;
\biggl\{\frac{9}{29},\frac{49}{29},\frac{67}{29},\frac{107}{29},
\cdots\biggr\}_{h=\frac{49}{29}}
\rightarrow\nonumber\\
&&\biggl\{\frac{4}{13},\frac{22}{13},\frac{30}{13},\frac{48}{13},
\cdots\biggr\}_{h=\frac{22}{13}}
\rightarrow\;\;
\biggl\{\frac{7}{23},\frac{39}{23},\frac{53}{23},\frac{85}{23},
\cdots\biggr\}_{h=\frac{39}{23}}
\rightarrow\;\;\;
\biggl\{\frac{3}{10},\frac{17}{10},\frac{23}{10},\frac{37}{10},
\cdots\biggr\}_{h=\frac{17}{10}}
\rightarrow\nonumber\\
&&\biggl\{\frac{5}{17},\frac{29}{17},\frac{39}{17},\frac{63}{17},
\cdots\biggr\}_{h=\frac{29}{17}}
\rightarrow\;\;
\biggl\{\frac{2}{7},\frac{12}{7},\frac{16}{7},\frac{26}{7},
\cdots\biggr\}_{h=\frac{12}{7}}\rightarrow\;\;\;\;\;
\biggl\{\frac{3}{11},\frac{19}{11},\frac{25}{11},\frac{41}{11},
\cdots\biggr\}_{h=\frac{19}{11}}
\rightarrow\nonumber\\
&&\biggl\{\frac{4}{15},\frac{26}{15},\frac{34}{15},\frac{56}{15},
\cdots\biggr\}_{h=\frac{26}{15}}
\rightarrow\;\;
\biggl\{\frac{5}{19},\frac{33}{19},\frac{43}{19},\frac{71}{19},
\cdots\biggr\}_{h=\frac{33}{19}}\rightarrow\;\;\;
\biggl\{\frac{6}{23},\frac{40}{23},\frac{52}{23},\frac{86}{23},
\cdots\biggr\}_{h=\frac{40}{23}}
\rightarrow\nonumber\\
&&\biggl\{\frac{1}{4},\frac{7}{4},\frac{9}{4},\frac{15}{4},
\cdots\biggr\}_{h=\frac{7}{4}}
\rightarrow\;\;\;\;\;\;\;\;\;
\biggl\{\frac{6}{25},\frac{44}{25},\frac{56}{25},\frac{94}{25},
\cdots\biggr\}_{h=\frac{44}{25}}\rightarrow\;\;\;
\biggl\{\frac{5}{21},\frac{37}{21},\frac{47}{21},\frac{79}{21},
\cdots\biggr\}_{h=\frac{37}{21}}
\rightarrow\nonumber\\
&&\biggl\{\frac{4}{17},\frac{30}{17},\frac{38}{17},\frac{64}{17},
\cdots\biggr\}_{h=\frac{30}{17}}
\rightarrow\;\;
\biggl\{\frac{3}{13},\frac{23}{13},\frac{29}{13},\frac{49}{13},
\cdots\biggr\}_{h=\frac{23}{13}}\rightarrow\;\;\;
\biggl\{\frac{2}{9},\frac{16}{9},\frac{20}{9},\frac{34}{9},
\cdots\biggr\}_{h=\frac{16}{9}}
\rightarrow\nonumber\\
&&\biggl\{\frac{5}{23},\frac{41}{23},\frac{51}{23},\frac{87}{23},
\cdots\biggr\}_{h=\frac{41}{23}}
\rightarrow\;\;
\biggl\{\frac{3}{14},\frac{25}{14},\frac{31}{14},\frac{53}{14},
\cdots\biggr\}_{h=\frac{25}{14}}\rightarrow\;\;\;
\biggl\{\frac{4}{19},\frac{34}{19},\frac{42}{19},\frac{72}{19},
\cdots\biggr\}_{h=\frac{34}{19}}
\rightarrow\nonumber\\
&&\biggl\{\frac{5}{24},\frac{43}{24},\frac{53}{24},\frac{91}{24},
\cdots\biggr\}_{h=\frac{43}{24}}
\rightarrow\;\;
\biggl\{\frac{6}{29},\frac{52}{29},\frac{64}{29},\frac{110}{29},
\cdots\biggr\}_{h=\frac{52}{29}}\rightarrow\;
\biggl\{\frac{1}{5},\frac{9}{5},\frac{11}{5},\frac{19}{5},
\cdots\biggr\}_{h=\frac{9}{5}}
\rightarrow\nonumber\\
&&\biggl\{\frac{6}{31},\frac{56}{31},\frac{68}{31},\frac{118}{31},
\cdots\biggr\}_{h=\frac{56}{31}}
\rightarrow\;
\biggl\{\frac{5}{26},\frac{47}{26},\frac{57}{26},\frac{99}{26},
\cdots\biggr\}_{h=\frac{47}{26}}\rightarrow\;\;
\biggl\{\frac{4}{21},\frac{38}{21},\frac{46}{21},\frac{80}{21},
\cdots\biggr\}_{h=\frac{38}{21}}
\rightarrow\nonumber\\
&&\biggl\{\frac{3}{16},\frac{29}{16},\frac{35}{16},\frac{61}{16},
\cdots\biggr\}_{h=\frac{29}{16}}
\rightarrow\;\;\;
\biggl\{\frac{5}{27},\frac{49}{27},\frac{59}{27},\frac{103}{27},
\cdots\biggr\}_{h=\frac{49}{27}}\rightarrow\;
\biggl\{\frac{2}{11},\frac{20}{11},\frac{24}{11},\frac{42}{11},
\cdots\biggr\}_{h=\frac{20}{11}}\rightarrow\nonumber\\
&&\biggl\{\frac{3}{17},\frac{31}{17},\frac{37}{17},\frac{65}{17},
\cdots\biggr\}_{h=\frac{31}{17}}
\rightarrow\;\;\;
\biggl\{\frac{4}{23},\frac{42}{23},\frac{50}{23},\frac{88}{23},
\cdots\biggr\}_{h=\frac{42}{23}}
\rightarrow\;\;
\biggl\{\frac{1}{6},\frac{11}{6},\frac{13}{6},\frac{23}{6},
\cdots\biggr\}_{h=\frac{11}{6}}\rightarrow\nonumber\\
&&\biggl\{\frac{3}{19},\frac{35}{19},\frac{41}{19},\frac{73}{19},
\cdots\biggr\}_{h=\frac{35}{19}}
\rightarrow\;\;\;
\biggl\{\frac{2}{13},\frac{24}{13},\frac{28}{13},\frac{50}{13},
\cdots\biggr\}_{h=\frac{24}{13}}
\rightarrow\;\;
\biggl\{\frac{1}{7},\frac{13}{7},\frac{15}{7},\frac{27}{7},
\cdots\biggr\}_{h=\frac{13}{7}}\rightarrow\nonumber\\
&&\biggl\{\frac{2}{15},\frac{28}{15},\frac{32}{15},\frac{58}{15},
\cdots\biggr\}_{h=\frac{28}{15}}
\rightarrow\;\;\;
\biggl\{\frac{1}{8},\frac{15}{8},\frac{17}{8},\frac{31}{8},
\cdots\biggr\}_{h=\frac{15}{8}}
\rightarrow\;\;\;\;
\biggl\{\frac{2}{17},\frac{32}{17},\frac{36}{17},\frac{66}{17},
\cdots\biggr\}_{h=\frac{32}{17}}\rightarrow\nonumber\\
&&\biggl\{\frac{1}{9},\frac{17}{9},\frac{19}{9},\frac{35}{9},
\cdots\biggr\}_{h=\frac{17}{9}}.\nonumber
\end{eqnarray}

\section{Final remarks}

We have verified that our approach to the FQHE 
reproduces all experimental data and can predicting 
the occurrence of this phenomenon 
for other filling factors. According to our formulation, the topological 
character of these quantum numbers is related with the Hausdorff 
dimension of the quantum paths of fractons. For that, we have obtained 
a physical analogous to the mathematical one formula 
of the Hausdorff dimension associated with fractal curves. The FQHE occurs in pairs 
of dual topological quantum numbers filling factors. The 
foundational basis of our theoretical formulation is 
free of any empirical formula and this characteristic 
constitutes the great difference between our insight 
and others of the literature. The filling factors are obtained from first-principles 
and the quantum Hall states are modelled  by fractons which carry rational 
or irrational values of spin and satisfy a fractal distribution function 
associated with a fractal von Neumann entropy. Also these objects satisfy a 
fractal-deformed Heisenberg algebra\cite{R8}. Therefore, the physical 
scenario takes place, i. e., the thermodynamical properties 
of such systems can be investigated.

We emphasize that our approach is supported by {\it symmetry principles} such as: 
mirror symmetry behind the fractal spectrum, duality symmetry 
between universal classes of particles, fractal supersymmetry, modular group behind 
the universality classes of the quantum Hall transitions. Besides 
these robust arguments given that the symmetry groups have 
great importance for understanding the physical theories even 
before any dynamics, we have established a connection between 
physics, fractal geometry and number theory. We have also obtained the results
\cite{R7,R8,R9,R10,R11,R12,R13,R14,R15}: a relation between the 
fractal parameter and the Rogers dilogarithm function, through the 
concept of fractal index, which is defined 
in terms of the partition function associated with each 
universal class of particles; a connection between the fractal 
parameter and the Farey 
sequences of rational numbers. Farey series $F_{n}$ of order 
$n$ is the increasing sequence of 
irreducible fractions in the range $0-1$ whose 
denominators do not exceed $n$. We have the following 

{\bf Theorem}\cite{R12}: {\it The elements of the Farey series $F_{n}$ 
of the order $n$, belong to the fractal sets, whose Hausdorff 
dimensions are the second fractions of the fractal sets. The 
Hausdorff dimension has values within the interval 
$1$$\;$$ < $$\;$$h$$\;$$ <$$\;$$ 2$ and these ones 
are associated with fractal curves.}

Along this discussion we have obtained fractal sets of dual 
topological quantum numbers filling factors associated with the FQHE. 
The universality classes of the 
quantum Hall transitions were established and the fractal 
geometry of nature in this context is manifest
in a simple and intuitive way. We believe that our formulation 
sheds some light on that phenomenon.

Finally, on the one hand, we know that the Farey sequences are 
related with the Riemann hypothesis\cite{R29} and 
on the other hand, we have established 
a connection between FQHE, fractal geometry and Farey sequences. So, 
we can conjecture if the FQHE is a convenient physical system 
for we take into account in a possible prove about the non-trivial 
zeros of the Riemann zeta function which states that they lie on the 
critical line ${\bf Re}(z)=1/2.$


\begin{thebibliography}{99}

\bibitem{R1} B. B. Mandelbrot, {\it The Fractal Geometry of Nature} 
(Freeman, New York, 1982);\\
C. Tricot, {\it Curves and Fractal Dimension} 
(Springer-Verlag, New York,1995);\\
K. Falkoner, {\it Fractal Geometry} (Wiley, New York, 1990);\\
K. Falkoner, {\it Techniques in Fractal Geometry} (Wiley, New York, 1997);\\
K. Falkoner, {\it The Geometry of Fractal Sets}
(Cambridge University Press, Cambridge, 1985).
\bibitem{R2} K. M. Kolwankar and A. D. Gangal, Pramana J. Phys. {\bf 48}, 49 (1997); 
Chaos {\bf 6}, 505 (1996);\\
A. Rocco and B. J. West,  Physica {\bf A265}, 535 (1999);\\
H. Kr\"oger, Phys. Rep. {\bf 323}, 818 (2000);\\
D. W\'ojcik {\it et al.}, Phys. Rev. Lett. {\bf B85}, 5022 (2000);\\
L. F. Abbott and M. B. Wise, Am. J. Phys. {\bf 49}, 37 (1981);\\
P. Meakin, {\it Fractals, Scaling and Growth far from Equilibrium}
(Cambridge University Press, Cambridge,1998);\\
R. P. Feynman and A. R. Hibbs, {\it Quantum Mechanics and Path Integrals} 
( MacGraw-Hill, New York, 1965 ), pp. 176-177.
\bibitem{R3}R. B. Laughlin, Rev. Mod. Phys. {\bf 71}, 863 (1999);\\
H. Stormer, Rev. Mod. Phys. {\bf 71}, 875 (1999);\\
D. C. Tsui, Rev. Mod. Phys. {\bf 71}, 891 (1999);\\
and references therein.
\bibitem{R4} D. J. Thouless, {\it Topological Quantum Numbers in Nonrelativistic Physics}  
( World Scientific, Singapore, 1998 );\\
F. Wilczek, cond-mat/0206122;\\
M. Stone, {\it The Quantum Hall Effect} ( World Scientific, Singapore, 1992 );\\
R. E. Prange and S. M. Girvin, {\it The Quantum Hall Effect}
(Springer Verlag, Heidelberg,1990);\\
J. E. Avron {\it et al.}, math-ph/0303055; J. E. Avron and L. Sadun, math-ph/0008040;\\
J. E. Avron and D. Osadchy, math-ph/0110026;\\
J. Bellissard {\it et al.}, J. Math. Phys. {\bf 35}, 5373 (1994).
\bibitem{R5} M. Marcolli and V. Mathai, math.DG/9803051; 
Commun. Comtemp. Math. {\bf 1}, 553 (1999); 
Commun. Math. Phys. {\bf 217}, 55 (2001);\\
A. L. Carey {\it et al.},  Commun. Math. Phys. {\bf 190}, 629 (1997); 
Lett. Math. Phys. {\bf 47}, 215 (1999); math.OA/0008115.
\bibitem{R6}S. Skoulakis and S. Thomas, Nucl. Phys. {\bf B538}, 659 (1999);\\
A. P. Balachandran {\it et al.},  Int. J. Mod. Phys. {\bf A11}, 3587 (1996);\\
Nucl. Phys. {\bf B443}, 465 (1995);\\
G. Murthy and R. Shankar, cond-mat/0205326;\\
 R. Shankar, cond-mat/0108271;\\
A. Comtet {\it et al.}, (Eds.) {\it Topological Aspects of Low Dimensional Systems, 
Les Houches 1998 }, {\bf vol. LXIX} ( Springer Verlag, New York, 1999 );\\
A. Khare, {\it Fractional Statistics and Quantum Theory} ( World Scientific, Singapore, 1998 );\\
G. Dunne, {\it Self-dual Chern-Simons Theories}(Springer Verlag, Heidelberg,1992);\\
A. Lerda, {\it Anyons: Quantum Mechanics of Particles with Fractional Statistics}
(Springer Verlag, Heidelberg,1995);\\
J. M. Leinaas, Rep. Phys. {\bf 242}, 371 (1994);\\
S. Forte, Rev. Mod. Phys. {\bf 64}, 193 (1992);\\
R. Iengo andK. Lechner, Rep. Phys. {\bf 213}, 179 (1992);\\
F. Wilczek, {\it Fractional Statistics and Anyon Superconductivity} 
( World Scientific, Singapore, 1990 );\\
F. Wilczek and Shapere, {\it Geometric Phases in Physics}
( World Scientific, Singapore, 1989 );\\
and references therein.
\bibitem{R7} W. da Cruz, Int. J. Mod. Phys. {\bf A15}, 3805 (2000).
\bibitem{R8} W. da Cruz, Physica {\bf A313}, 446 (2002).
\bibitem{R9} W. da Cruz and R. de Oliveira, Mod. Phys. Lett. {\bf A15}, 1931 (2000).
\bibitem{R10} W. da Cruz, J. Phys: Cond. Matter. {\bf 12}, L673 (2000).
\bibitem{R11} W. da Cruz, Mod. Phys. Lett. {\bf A14}, 1933 (1999).
\bibitem{R12} W. da Cruz, Chaos, Solitons and Fractals {\bf 17}, 975 (2003).
\bibitem{R13} W. da Cruz, Int. J. Mod. Phys. {\bf A18}, 2213 (2003). 
{\it Proceedings 2nd International Londrina Winter School: 
Mathematical Methods in Physics, August, 26-30, 2002}, 
Eds. M. C. B. Abdalla {\it et al.}
\bibitem{R14} W. da Cruz, cond-mat/0301587.
\bibitem{R15} W. da Cruz, cond-mat/0304398.
\bibitem{R16} A. M. Polyakov, in {\it Proc. Les
 Houches Summer School
 {\bf vol. IL}}, ed. E. Br\'ezin and J. Zinn-Justin
  (North Holland, 1990), p.305.
\bibitem{R17} F. D. Haldane, Phys. Rev. Lett. {\bf 67}, 937 (1991);\\
Y. S. Wu, Phys. Rev. Lett. {\bf 73}, 922 (1994);\\
S. B. Isakov, Mod. Phys. Lett. {\bf B8}, 319 (1994);\\ 
A. K. Rajagopal, Phys. Rev. Lett. {\bf 74}, 1048 (1995);\\ 
A. D. de Veigy and S. Ouvry, Phys. Rev. Lett. {\bf 72}, 600 (1994).
\bibitem{R18} A. Kirillov, Prog. Theor. Phys. Suppl. {\bf 118}, 61 (1995).
\bibitem{R19} R. G. Mani and K. von Klitzing, Z. Phys. {\bf B100}, 635 (1996).
\bibitem{R20} J. H. Smet, Nature {\bf 422}, 391 (2003).
\bibitem{R21} A. W\"ojs, K. S. Yi and J. J. Quinn, cond-mat/0304130.
\bibitem{R22} C. C. Chang {\it et al.}, Phys. Rev. {\bf B67}, 121305(R) (2003).
\bibitem{R23} W. Pan {\it et al.}, Phys. Rev. Lett. {\bf 90}, 016801 (2003).
\bibitem{R24} W. Pan {\it et al.}, Phys. Rev. Lett. {\bf 88}, 176802-1 (2002).
\bibitem{R25} R. R. Du {\it et al.}, Phys. Rev. Lett. {\bf 75}, 3926 (1995).
\bibitem{R26} R. R. Du {\it et al.}, Phys. Rev. Lett. {\bf 70}, 2944 (1993).
\bibitem{R27} R. Willett {\it et al.}, Phys. Rev. Lett. {\bf 59}, 1776 (1987).
\bibitem{R28} B. P. Dolan, J. Phys. {\bf A32}, L243 (1999); 
Nucl. Phys. {\bf B554}, 487 (1999);\\
C. P. Burgess and C. A. L\"utken, Nucl. Phys. {\bf B500}, 367 (1997);\\
and references therein.
\bibitem{R29} J. Franel, G\"ottinger Nachrichten, pp. 198-201 (1924);\\
 E. Landau, G\"ottinger Nachrichten, pp. 202-206 (1924);\\
H. M. Edwards, {\it Riemann's Zeta Function} ( Dover, New York, 1974 ), pp. 263-267;\\
J. B. Conrey, Notices of the AMS. {\bf 50}, 341 (2003);\\
M. R. Watkins, http://www.maths.ex.ac.uk/$\sim$mwatkins/zeta/physics.htm.
\end{thebibliography}
\end{document}